\documentclass[a4paper]{llncs}

\usepackage{amssymb}
\usepackage{graphicx}
\usepackage{url}

\newcommand{\keywords}[1]{\par\addvspace\baselineskip
\noindent\keywordname\enspace\ignorespaces#1}

\newcommand{\code}[1]{{\ttfamily #1}}
\newcommand{\refactoring}[1]{\textsc{#1}}
\newcommand{\pattern}[1]{\textsc{#1}}

\begin{document}

\mainmatter  

\title{Clone Removal in Java Programs \\
as a Process of Stepwise Unification}

\author{Daniel Speicher \and Andri Bremm}

\institute{University of Bonn, Computer Science III,\\
R\"omerstra\ss{}e 164, 53117 Bonn, Germany\\
dsp@cs.uni-bonn.de, bremm@uni-bonn.de}

\maketitle

\begin{abstract}
Cloned code is one of the most important obstacles against consistent software
maintenance and evolution. Although today's clone detection tools find a variety
of clones, they do not offer any advice how to remove such clones. We explain
the problems involved in finding a sequence of changes for clone removal and
suggest to view this problem as a process of stepwise unification of the clone
instances. Consequently the problem can be solved by backtracking over the
possible unification steps.
\end{abstract}

\keywords{unification, backtracking, clone analysis, refactoring, program
dependence graph, lambda expression}

\section{Introduction}

In the last decades, practicable and reliable code clone detection
tools~\cite{Roy09} have been developed. These tools are able to find different
kinds of clones, which are all important in some situations. Developers, who are
interested in the removal of clones by refactoring~\cite{Fowler99}, want to know
whether and how a clone can be eliminated. A refactoring is a change to the
source code that alters (typically improves) the design, but does not change the
observable behavior of the software. We present an approach that gives precise
refactoring suggestions depending on the set of available refactorings.

Fowler et al. provided in \cite{Fowler99} a comprehensive catalog of such
refactorings. A developer may use for example the \refactoring{rename method}
refactoring to change the name of a method to a more expressive one. A tool that
offers this refactoring has to make sure that the name of the method is not only
changed in the declaration of the method, but as well at every method
invocation. As the same name might be used for different methods in different
scopes, a textual "search and replace" is not guaranteed to keep the observable
behavior intact. The tool needs to know the Abstract Syntax Tree of the code
including resolved bindings to methods. JTransformer~\cite{JTransformer}
provides this information for Java programs as facts on which logic programs can
reason.

{\em Identical} clones can be removed by \refactoring{extract method},
\refactoring{extract class}, \refactoring{pull up method} refactorings together
with the appropriate adaptations at the call side. If the clones have some {\em
differences}, we suggest to start with the Program Dependence
Graphs~\cite{Ferrante87} (PDG) for the clone instances, to identify the
statements that are {\em equal or unifiable} (i.e. can be made equal through
refactorings) and finally rearrange the control flow based on the PDG so that
all non-unifiable statements are separated. The example in the Figures~1 to 3
illustrates our approach.

\begin{figure}[!t]

\includegraphics[width=3.5in]{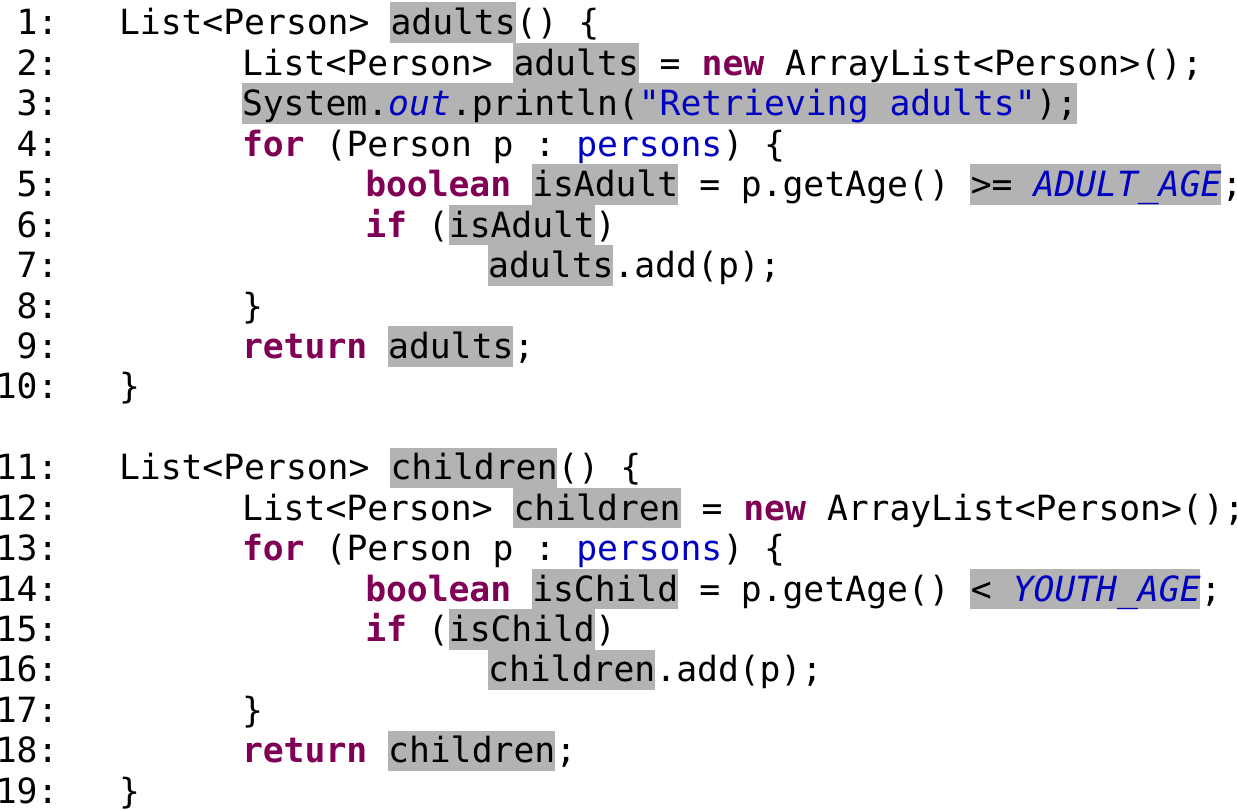}

\caption{Two cloned methods. The methods differ in the name of two
variables (\code{adults} vs. \code{children}, \code{isAdult} vs.
\code{isChild}), one extra statement (line 3), and a non-unifiable expression
(line 5 vs. line 14).}
\label{fig_clones}
\end{figure}

\section{Program Dependence Graphs}

We start with two potential clone candidates. These may have been found with one
of the existing clone detection tools like Simian \cite{Simian} or Scorpio
\cite{Higo11,Scorpio}. For each of the clone candidates we build the PDG.

Such a PDG consists of one node for every statement and of two kind of edges
representing control and data dependencies.  There is a {\em control dependency}
from a control statement to all directly enclosed statements. In our example the
for-loop in line 4 controls the execution of line 5 and 6 while the execution of
line 7 in turn is controlled by line 6. There is a {\em data dependency} from a
statement $s_1$ to a statement $s_2$, if $s_1$ writes a variable that is read in
$s_2$ and there is at least one possible execution  on which $s_1$ is the last
statement writing this variable before reaching $s_2$. In our example there is a
data dependency from statement 2 to statement 9 as the for-loop might not be
executed.

\begin{figure}[t]
\centering

\includegraphics[width=3.7in]{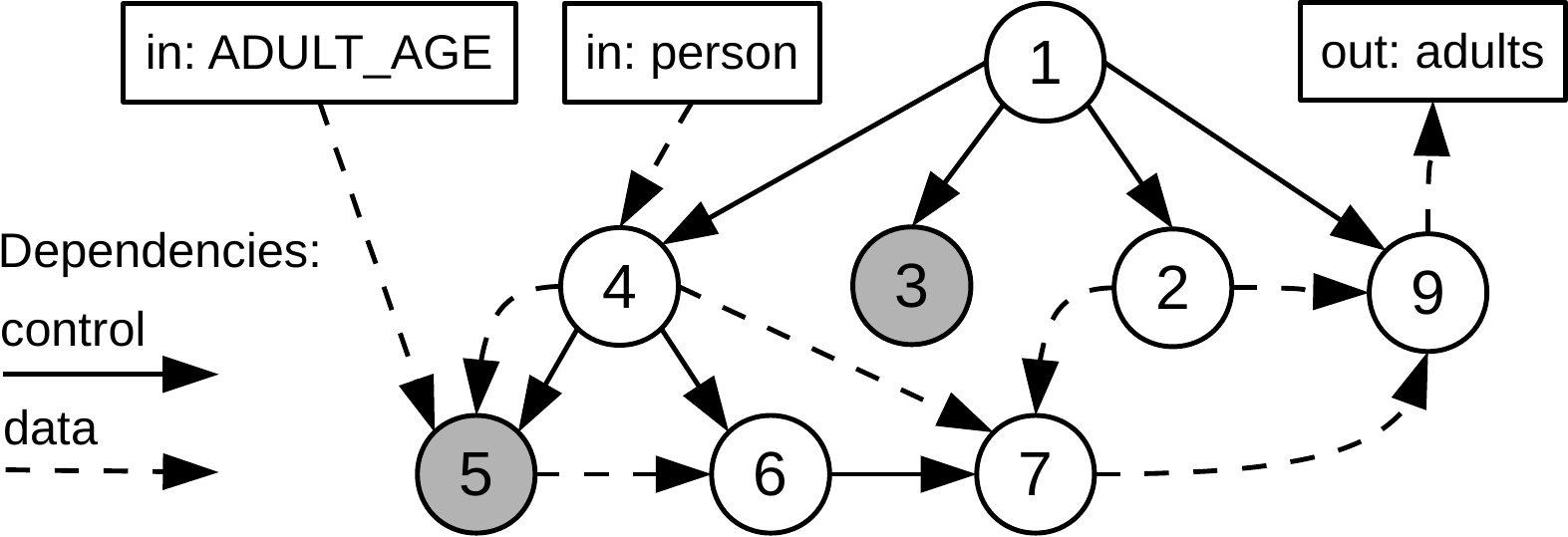}

\caption{A PDG of the method \code{adults} in Figure \ref{fig_clones}. The
invocation of \code{add} in line~7 is interpreted as a write on \code{adults}
leading to the data dependency from 7 to 9. The invocation of \code{getAge} in
line~5 is interpreted as a read on \code{p}, so that there is no data dependency
from 5 to 7. The extra statement in line~3 has no data dependencies, so that
it can freely be moved below node~1. The data dependencies of the non-unifiable
statement in line~5 forbid reordering but allow extracting into a lambda
expression.}
\label{fig_pdg}
\end{figure}

In extension to the established definition our data dependencies take as well method
invocations into account. If a method returns a value without performing side
effects we consider the method invocation only as a {\em read access} to the object.
If the method does have side effects, we consider the invocation as a {\em write
and read access} to the object. The PDG in Figure \ref{fig_pdg} gives an example
of a data dependency resulting from this approach.
This is still a heuristic way to transfer the concept of a data dependency to
the object-oriented setting. Deeper analysis could label the data dependencies
with a more precise characterization of the state that is changed by the method.
In addition alias analysis could find additional hidden dependencies as a change
to one object via one variable is a change to the object behind its aliases.

Once we have build the PDGs of the clone candidates, they are compared and nodes
for equal or unifiable statements are mapped to each other. Whether two
statements are unifiable depends on the refactorings that are considered.

\section{Statement Unification}

\begin{figure}[!t]

\includegraphics[width=3.8in]{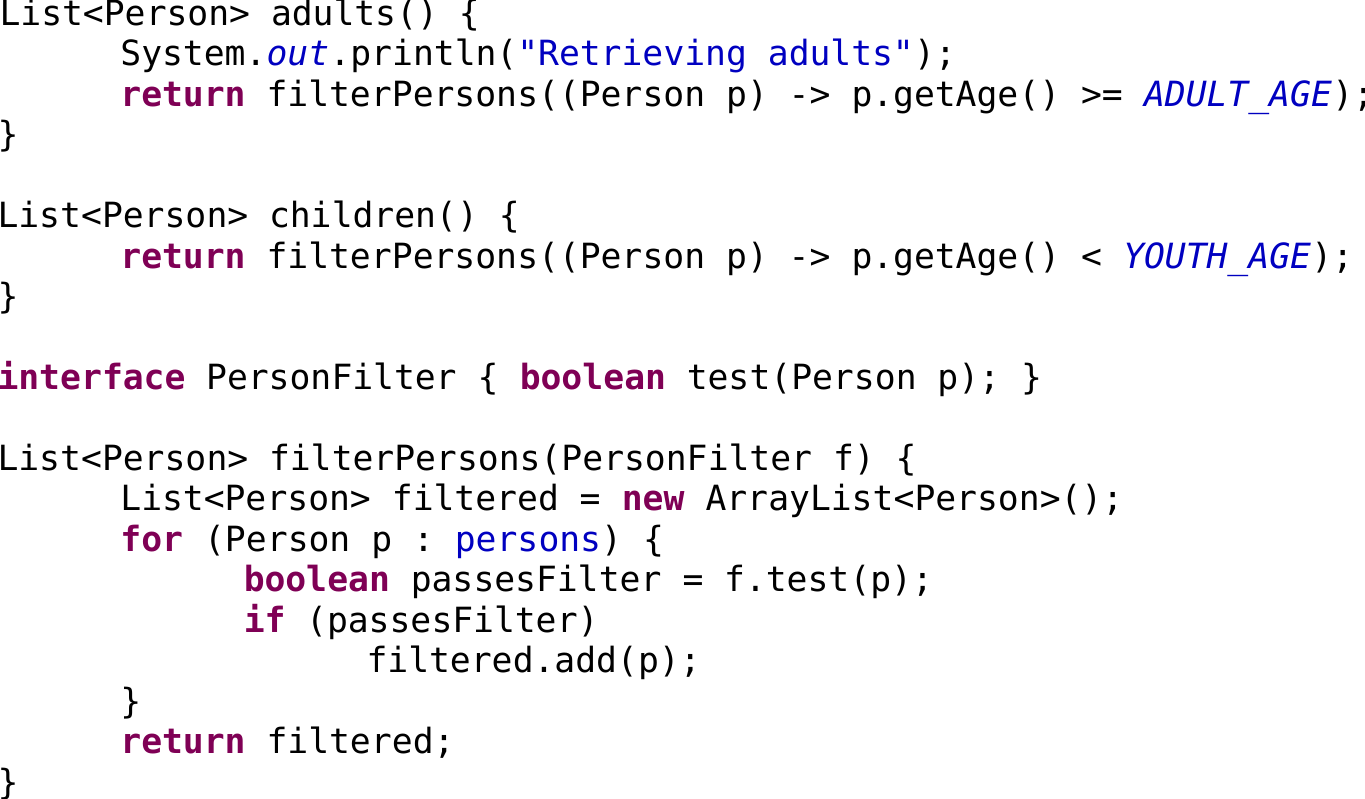}

\caption{The methods after the clone refactoring. The differing variables were
renamed. The extra statement in the first method has been separated from the
contiguous block of unifiable statements. The non-unifiable expressions have
been extracted into lambda expressions. Lambda expressions consist of a
parameter list surrounded by round brackets. The statements of the lambda
expression follow after the arrow. If there is only one statement it is
possible to omit the return keyword.}
\label{fig_refactored}
\end{figure}

The \refactoring{rename} refactoring allows us to consistently change the names
of local variables and parameters or even fields and methods. Therefore we take
at least the \refactoring{rename} refactoring into account. Our example in
Figure~\ref{fig_clones} and \ref{fig_refactored} illustrates this. If we
consider further refactorings more statements become unifiable although some
at the price of complexer parameter lists.

Differences in literals can be removed with the \refactoring{introduce
parameter} refactoring. The \refactoring{generalize types} refactoring
\cite{Tip03} allows to find differences in type declarations that are more
specific than required by the usage of the declared object. If a type
generalization is not possible e.g. because different specific return types are
required by the callers of a method, the refactoring \refactoring{introduce type
parameter} can help. Finally method signatures that differ just in the order of
parameters can be unified with the \refactoring{reorder parameters} refactoring.

In our example the difference between line~5 and line~14 can not be removed and
the statement in line~3 has no counterpart in the second method. These
differences require changes to the control flow.


\section{Control Flow Unification}

The PDG contains only as much information about the control flow as is relevant
for the state of the variables at each line of the method. Therefore a node can
freely be reordered (directly) below the node that controls it as long as the
order of nodes with data dependencies is preserved.\footnote{Tsantalis and
Chatzigeorgiou in \cite{Tsantalis09} correctly emphasize that there is one
additional criterion to be considered.
Although a write access to a variable {\em following} a read access can not
influence the value of the variable, and therefore is not represented by a data
dependency, moving the write access upwards may create a data dependency and
change the behavior. Adding these so called "anti-dependencies" to the PDG an
preserving their order solves the problem. Our example does not show
anti-dependencies.} This allows us to separate non-unifiable statement from
unifiable statements, as it is the case with node~3 in our example.

Another possibility to ``remove'' non-unifiable statements is to use the
\refactoring{extract method} refactoring. A group of contiguous statements is
extractable if the corresponding nodes have to other statements only outgoing
data dependencies for at most one variable and no outgoing control
dependency~\cite{Tsantalis09}. In the PDG in Figure~\ref{fig_pdg} the nodes~5,
6, and 7 together as well as the node~5 on its own is extractable.

The \refactoring{extract method} refactoring is especially helpful if the clones
are in classes that are siblings in the class hierarchy. If in this case all
differences can be extracted the remaining method can be \refactoring{pulled up}
to a common ancestor of the siblings. This sequence of refactorings is called
\refactoring{form template method} and is explained in detail in \cite{Fowler99}.

If the classes containing the clones are unrelated the \pattern{strategy} design
pattern in combination with \pattern{template method} may be used
\cite{Balazinska00}. But, if there is only one or two differences and these
differences are small, these pattern do not pull their weight and the
introduction of lambda expressions is the method of choice.\footnote{Lambda
expressions are essential for every functional language and have been available
for some object-oriented languages as well. Finally they will be introduced to
Java in the next version. The planed syntax for lambda expression in Java is
explained in JSR~335~\cite{Goetz12}. } The preconditions for the
\refactoring{extract lambda expression} refactoring are the same as for
\refactoring{extract method}. Our example illustrates the use of lambda
expressions to extract the difference between line~5 and 14.

\section{Related Work and Conclusion}

CloneDifferentiator~\cite{Xing11} analyses and visualises the differences
between the PDG of clones. The refactorings \refactoring{extract method},
\refactoring{introduce parameter} and the use of Generics are suggested.
ARIES~\cite{Higo05} calculates metrics to decide whether a refactoring is
appropriate. For example \refactoring{extract method} is only recommended when
the fragment refers to only a few variables outside the fragment.

We described a process that derives for related clones one (or more) ways to
remove the clones, by applying a series of refactorings. The parameters of the
refactorings can be precisely (although not necessarily uniquely) derived from
the context so that a tool can present precise refactoring suggestions to the
developer. Elements of the presented approach such as the generation of the PDG
are implemented as part of Cultivate \cite{Cultivate}.

The approach to search for a sequence of refactoring steps by exploring the
different possibilities to unify statements and control flow naturally arises
from the problem. As we start with clone candidates found by existing tools, the
amount of data to be processed is limited: We know already which methods to
compare and do not have to compare all possible method pairs. In addition
typically only a few statements in the methods are unifiable, so that the graph
matching is not as expensive as in the general case.


\def\IEEEbibitemsep{2pt plus .5pt}

\end{document}